# Facet-dependent Chemical Kinetics Governed Growth of Twisted Graphene Layers with Pre-designed Angles

Chaowu Xue[1,2#], Mengzhao Sun[1,2#], Zixuan Zhou[1#], Zhuoran Yao[1], Li-Qun Shen[1], Xiao Kong[3]*, Honglong Zhao[1], Feng Ding[4]*, Marc Willinger[5], Zhongkai Liu[1]*, Zhu-Jun Wang[1]*

[1]School of Physical Science and Technology, ShanghaiTech University, Shanghai, China.
[2]Center for Transformative Science, Shanghai Tech University, Shanghai, China.
[3]Shanghai Institute of Microsystem and Information Technology, Chinese Academy of Sciences, Shanghai, China.
[4]Suzhou Laboratory, Suzhou, China.
[5]Department of Chemistry, Technical University of Munich, Garching, Germany.

[#]These authors contributed equally: Chaowu Xue, Mengzhao Sun, Zixuan Zhou.
*Corresponding authors: xkong91@mail.sim.ac.cn; dingf@szlab.ac.cn; liuzhk@shanghaitech.edu.cn; wangzhj3@shanghaitech.edu.cn.

**Abstract:**

Twisted graphene layers (TGLs) provide a powerful platform for investigating multiple quantum phenomena, yet their scalable deployment is hindered by the lack of reliable synthesis with precise angle. Here, benefited from a deeper understanding of the interplay between grain index and graphene growth kinetics, we report a scalable strategy to grow TGLs with pre-designed twist angles on platinum (Pt) via chemical vapor deposition (CVD), Through a combination of complementary *in situ* methods, we identified the activity sequence of different Pt grains and attributed it to the area ratio of exposed (110) facets during graphene-induced surface reconstruction. Moreover, we revealed that CVD-grown graphene orientation is determined by the grain-orientation-dependent surface morphology. By leveraging the so-established correlations between grain index with both graphene growth priority and its orientation, we achieve controlled folding and tearing of graphene overlayer using a pair of adjacent grains with dramatically different catalytical activity and kink-free atomic steps. We reveal that overlayer-induced step bunching and terrace reconfiguration critically govern the domain morphology and folding direction. Building on this mechanistic insight, we demonstrate a substrate-engineering framework where specific platinum grains are rationally selected to yield TGLs with pre-designed twist angles, including magic angle with flat band dispersion. This work not only highlights fundamental kinetics of Pt catalyzed graphene CVD growth, but also offers a generalizable methodology for manipulating foldable two-dimensional materials via dynamic substrate reconstruction, exampled by programmable growth of high-quality TGLs on open surfaces.

Key words: 2D materials, chemical vapor deposition, twisted graphene layers, twist-angle engineering, facet-dependent catalysis, multiscale *in situ* characterization

TGLs offer a transformative platform for exploring emergent novel physics, such as unconventional superconductivity, orbital magnetic moments, symmetry-broken states, and nontrivial topological phases[1-10]. These unique properties have inspired numerous prototypes for next-generation electronic and optical devices[11-14]. However, realizing the practical potential of TGLs requires a scalable method to synthesize large-area bilayers or multilayers with precise and homogeneous twist angles and pristine interfaces[15-17]. Current approaches to fabricating TGLs predominantly rely on post-growth manipulation techniques, such as polymer-assisted mechanical tearing and stacking[18-22]. Despite tunability to a certain extent, these methods suffer from critical limitations—including low scalability, poor precision, and interlayer contamination—rendering them impractical for device integration.

CVD provides a highly scalable alternative for two-dimensional (2D) material synthesis, yet direct growth of twisted 2D material on epitaxial substrates often fail to escape the energetic trap of AB-stacking or suffer from random nucleation at defects. To achieve programmable twists, replication methods using pre-rotated metal foils (forming a Cu/TGL/Cu sandwich structure)[20] have been attempted, but these involve complex wet-etching processes that introduce contamination and exhibit low yield. Moreover, only bilayer graphene can be achieved in this framework, which limits the degrees of freedom over the layer number and stacking order. Consequently, a strategy for direct, controllable growth of TGLs on an open surface remains elusive.

Among transition metal catalysts, Pt is uniquely suited for multi-layer graphene synthesis due to its high catalytic activity of hydrocarbon dissociation, ability to support single-crystalline overlayer growth, and relatively weak interaction with overlayers that facilitates transfer to arbitrary substrates[23-28]. However, harnessing Pt's full potential requires moving beyond a static view of the substrate behaviors during overlayer growth[29, 30]. It is crucial to develop a quantitative understanding of how the Pt surface dynamically evolves under reaction conditions—specifically, how facet-dependent catalytic activity and overlayer-induced surface reconstruction govern the nucleation and structure evolution (such as folding and tearing) of graphene[23, 25, 31-34]. This requires a comprehensive understanding as well as solid *in situ* characterization evidence of the chemical kinetics during graphene growth on general Pt grains.

To address the lack of efficient strategy in controlling the twist-angle of overlayers, we systematically investigated the thermodynamics and kinetics of graphene formation on polycrystalline Pt using a suite of *in situ* techniques (environmental scanning electron microscopy (ESEM), X-ray photoelectron spectroscopy (XPS), electron backscatter diffraction (EBSD), atomic force microscopy (AFM), and scanning tunneling microscopy (STM)). By tracking the evolution of individual Pt grains and correlate crystallographic orientations with catalytic kinetics, we mapped the catalytic activity of distinct Pt facets, revealing an unexpected activity order of {111} > {110} > {100} for precursor dissociation and graphene formation. These findings challenge the conventional view in heterogeneous catalysis that a high density of active sites and open surface structures are necessarily superior. Our observations reveal that the graphene–metal interaction is highly dynamic: graphene growth induces massive surface step bunching and reconstruction, amplifying the interfacial area. Furthermore, this geometric evolution, driven by the migration of grain boundaries (GBs) between facets of differing activity, triggers a deterministic folding and tearing process[35].

Building on these mechanistic and chemical insights, we established a facet-dependent engineering framework based on constrained growth and folding of graphene at Pt GB to deterministically set the twist angle of overlayers. This strategy integrates three cooperative effects: 1. Facet-dependent catalytic activity: leveraging the activity order to confine graphene nucleation to the high active "nucleation facet"; 2. Reconstruction-driven folding: utilizing the interfacial area amplification factor ($f_{IAF}$) on a neighboring "reconstruction facet" with lower activity but high capacity to drive folding via GB migration; 3. Step-edge orientation control: using close-packed <110> surface step edges to lock both overlayer in-plane lattice orientation and its folding axis.

By rationally pairing Pt grains with specific surface orientations to form artificial bi-crystals, we realize CVD synthesis of TGLs with pre-designed twist angles on open catalyst surfaces. Angle-resolved photoemission spectroscopy (ARPES) measurements confirm that the resulting twisted layers exhibit the predicted electronic band features, such as flat bands (for magic angle) and separated Dirac cones. This combination of atomic-scale insight and macroscopic substrate engineering provides a scalable, deterministic route for producing programmable twisted graphene, establishing the manufacturing foundation of scalable twistronic device.

**Results and discussion**

To elucidate how Pt crystallography dictates the nucleation, reconstruction, and folding processes that control graphene twist angles, we structured our investigation to logically progress from intrinsic polycrystalline Pt surface properties to dynamic overlayer growth behaviors (**Figures 1–6**). We begin with quantifying the catalytic activity of distinct Pt facets via *in situ* ESEM–EBSD correlation and the combination of density-functional theory (DFT) and kinetic Monte Carlo simulations, establishing a facet-specific activity order that governs the initial nucleation landscape (**Figure 1**). Multimodal ESEM, AFM, STM, and XPS analyses subsequently reveal that graphene–metal interactions drive pronounced surface reconstruction and step bunching with interfacial area amplified, which is essential for controlled folding (**Figure 2**). Integrating these thermodynamic and kinetic foundations, we subsequently directly visualize, through *in situ* ESEM, the dynamic evolution of graphene overlayers across neighboring Pt grains, capturing how the interplay of differential activity and reconstruction triggers spill-over graphene growth and facet-dependent folding and tearing (**Figure 3**). Building on these insights, we show how the geometry of close-packed <110> step edges governs in-plane graphene orientation and wrinkle direction, linking atomic-scale step structure to macroscopic interlayer twist (**Figure 4**). Integrating catalytic-activity order, interface reconstruction and step-edge orientation dependent folding, we construct a predictive design map that enables deterministic facet pairing for preset twist angles (**Figure 5**). Finally, we experimentally implement this strategy using an Orientation-Controlled Surface Polisher (OCSP), producing bi-crystal Pt substrates with programmed facet pairs (one "nucleation facet" for nucleation and one "reconstruction facet" for folding) and validating the resulting TGLs through AFM and ARPES measurements that confirm the designed interlayer angles and corresponding electronic structures (**Figure 6**).

**Grain-dependent catalytic activity of Pt for graphene growth**

To quantitatively link the growth dynamics to crystallography, we first examine how catalytic activity varies with Pt grain orientation using *in situ* ESEM-EBSD correlation (setup and schematic in **Figure 1a,b**). Under stepwise increase in hydrocarbon partial pressure, we continuously recorded facet-specific nucleation and growth (**Figure 1c-f**), thereby establishing a surface structure-activity relationship between facet index and catalytic performance. Direct observations reveal a pronounced induction period before graphene appears, after which nucleation and expansion occur selectively on facets of specific orientation (**Figure 1g**) while other facets remain bare under the same hydrocarbon partial pressure and exposure time. The order in which surfaces get covered as pressure rises shows evidently that Pt catalytic activity depends heavily on crystallographic orientation.

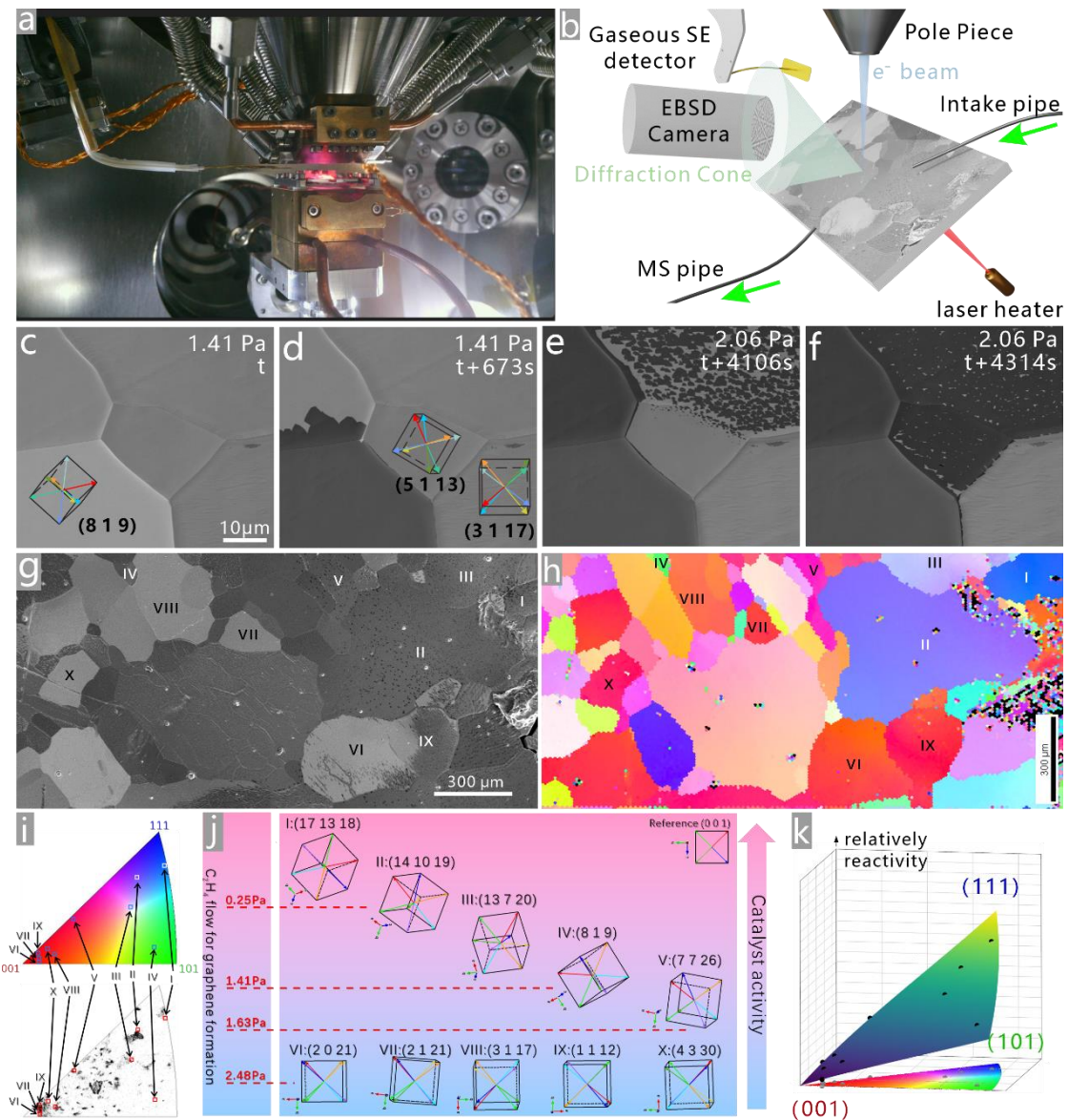

**Figure 1.** Catalytic activity ordering of Pt grain orientations. (a) *In situ* ESEM physical distribution under working conditions. (b) Schematic model of *in situ* ESEM operation; green arrows indicate gas flow direction. (c, d) Graphene growth on Pt foil under 1.41 Pa $C_2H_4$ and 23.59 Pa $H_2$ for 3788 s. (e, f) Graphene growth under 2.06 Pa $C_2H_4$ and 22.94 Pa $H_2$. Cubes in (c, d) represent grain orientations and surface indices. (g) ESEM image of polycrystalline Pt foil partially covered by graphene (2.25 Pa $C_2H_4$, 25 Pa $H_2$, 1000 °C, 4307 s). (h) Stereographic grain orientation map of the region in (g). (i) Pole figure of grains from (g) and (h); black arrows indicate positions and color legends for grains labeled with Roman numerals. (j) Relative catalytic activity of graphene growth on Pt grains ranked by $C_2H_4$ partial pressure. Orientation cubes represent spatial and angular orientations relative to the (001) reference. (k) Simulated Pt surface reactivity for graphene growth; black dots correspond to surfaces in (g–j).

EBSD on the same field of view (**Figure 1g,h**) allowed us to index each Pt facet in the inverse pole figure (IPF, **Figure 1i**) and to pointwise map the ethylene ($C_2H_4$) partial-pressure threshold for nucleation onto the corresponding Miller indices within the cubic orientation frame (**Figure 1j**),[36] from which the activity order of all crystal faces was determined. This three-factor mapping—real-space imaging, Pt facet orientation, and hydrocarbon pressure threshold—reveals a clear trend: at a given pressure, {111}-type orientations nucleate most readily and grow rapidly, followed by facets with substantial {110}-type character, whereas {100}-type orientations are the most difficult for graphene to cover.

To rationalize these facet-dependent differences, we develop a high-index surface decomposition strategy and combine DFT to quantify the dissociation pathways of representative precursors together with the Pt surface reconstruction induced by overlayer covering (**Figure 1k**). The results yield an activity order of {111} > {110} > {100}, in excellent agreement with *in situ* observation. This delineates a coherent picture in which the catalytic activity for graphene formation varies systematically with Pt facet orientation. More importantly, these findings challenge the conventional view in heterogeneous catalysis that a high density of active sites and open surface structures are necessarily superior[37, 38], indicating that the overlayer–catalyst interaction and resulting surface reconstruction play a decisive role in graphene formation on Pt.

## Graphene-induced substrate surface reconstruction

With the facet-dependent activity order established, we investigate how graphene growth modifies the underlying Pt surfaces. The interaction drives reconstruction and step bunching that reshape the graphene-metal interface, generating facet-dependent differences in the overlayer-substrate interfacial area[39]. *In situ* ESEM and quasi-*in situ* AFM show that on high-index facets far away from low-index orientations (*e.g.* Pt(7 7 26), **Figure 2a,b**), graphene growth induces aggregation of monoatomic steps into pronounced step bunches[40-45]; by contrast, on facets closer to low-index orientations (*e.g.* Pt(2 1 21)), no comparable bunching is observed upon graphene coverage. Using EBSD-determined facet orientations and surface atomic structures, we establish a correspondence from lattice orientation to atomic-scale step structure and eventually to reconstructed surface morphology, demonstrating that meso-scale reconstruction is governed by atomic-scale step bunching.

To understand the underlying mechanism, we performed quasi-*in situ* STM to directly probe the interplay between graphene nucleation and Pt surface. On Pt(34 35 34), a vicinal surface miscut by ~0.8° from (111) with moderate step density (**Figure 2c**), STM reveals that graphene nucleates preferentially at surface steps (**Figure 2d**; White dashed lines denote the initial step positions, while green and yellow arrows indicate the stepped surfaces that have extended following the growth of graphene.). As growth proceeds, steps collectively migrate and retreat (**Figure 2d,g**; green/yellow arrows denote uphill/downhill motion), transforming the original high density of monoatomic steps into wide, multi-atom-high steps and broader (111) terraces (positions ①-③ in **Figure 2g** showing the locations of the step edges that expand as the graphene grows) while preserving clear step-bunching morphology[40](**Figure 2e,f**). Concurrently, growth perpendicular to the step direction slows markedly, distorting the initially six-fold symmetric front into an anisotropic shape elongated along the steps (**Figure 2g**). These observations support a sequential pathway: step-edge nucleation, step migration/extension and step bunching.

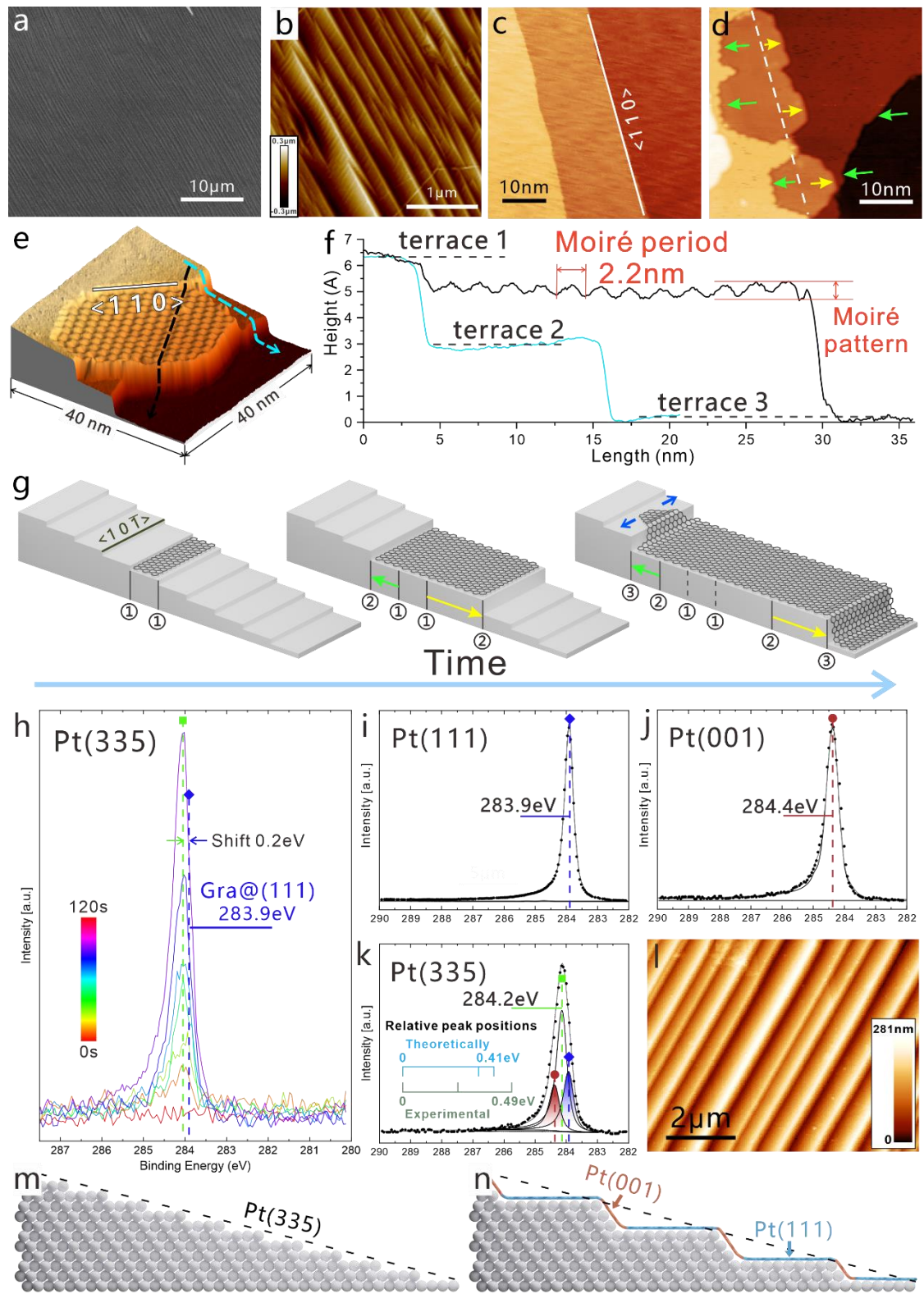

**Figure 2.** Graphene-induced step bunching and surface reconstruction. (a) ESEM and (b) high-resolution AFM images of graphene fully covering the Pt(7 7 26) surface, showing induced surface reconstruction. (c) STM image of a clean Pt(111) surface revealing relatively straight steps. (d, e) STM images after graphene growth illustrating step-edge retraction and terrace expansion. White dashed lines in (d) indicate step

positions before graphene formation (marked as white lines in (c) and (e)). (f) Height profile along the blue and black dashed arrows in (e); red arrows mark the height difference and period of the Moiré pattern. (g) Graphene growth models illustrating terrace extension, step bunching, and spill-over processes. Green arrows in (d, g) indicate Pt step-edge retraction, yellow arrows mark expansion, and blue arrows in (g) indicate the direction of graphene lateral growth. Numbers in (g) denote terrace-edge positions during growth evolution. (h) *In situ* XPS data of graphene growth on the Pt(335) surface over 120 s. Green and blue arrows indicate signal peak positions of graphene on the Pt(111) plane and complete coverage on the Pt(335) surface. (i–k) XPS spectra highlighting **C1s** signals for graphene on Pt(111), Pt(001), and Pt(335) surfaces; red, blue, and green dashed lines mark characteristic peak positions. (l) AFM image of graphene on Pt(335) showing step-bunched morphology. (m, n) Side-view atomic models of bare Pt(335) and graphene-covered Pt(335) surfaces.

To correlate morphological evolution with electronic/chemical changes at larger scales, we performed *in situ* near-ambient-pressure X-ray photoelectron spectroscopy (NAP XPS) on Pt(335), dynamically tracking the chemical shift of C 1s peak as a fingerprint of graphene–Pt coupling strength (**Figure 2h**) which directly reflects local bonding and reconstruction-induced electronic state changes.[46, 47] During early graphene growth on Pt(335), the C 1s peak appears at 283.9 eV (The blue arrow in **Figure 2h** indicates the location); with increasing coverage, the intensity rises and the peak shifts toward ~284.1 eV (The green arrow in **Figure 2h** indicates the location) . Comparison with static XPS data on standard low-index facets that C 1s peak is at 283.9 eV on Pt(111) and 284.4 eV on Pt(001) (**Figure 2i–k**), the broader C 1s envelope on Pt(335) can be attributed into contributions from graphene bound on (111)- and (001)-type surfaces through peak splitting analysis (**Figure 2k**). This reflects the reconstruction kinetics: initial nucleation on (111) terraces (peak near 283.9 eV) followed by a shift toward ~284.1 eV as graphene covers reconstruction-generated (001) facets—consistent with the step-faceted morphology probed by AFM (**Figure 2l**). Moreover, DFT results indicate a graphene binding energy difference of ~0.41 eV between Pt(111) and Pt(001), which is close to the experimental peak separation (~0.49 eV) and thus further supporting the linkage between surface reconstruction and C 1s peak shift (see **Figure 2k** inset).

Integrating *in situ* XPS, AFM, and STM with theory across diverse scales (**Figure 2h-k**), we conclude that graphene growth proceeds predominantly via step migration culminating in meso-scale step bunching. The ensuing reconstruction markedly enhances interfacial corrugation and thus amplifies the actual graphene-substrate contact area. To quantify this geometric amplification, we define the interfacial area amplification factor

$$f_{\text{IAF}} = \frac{A_{\text{int}}}{A_{\text{proj}}} \geq 1, \tag{1}$$

where $A_{\text{int}}$ is the reconstructed interfacial area and $A_{\text{proj}}$ is the projected (initial) area (Eq. 1). For Pt(335) (**Figure 2m**), after graphene-induced formation of periodic alternating (001) and (111) facets, the actual and initial areas (solid and dashed outlines in **Figure 2n**) yield $f_{\text{IAF(335)}} \approx 1.10$. Crucially, this graphene-induced step migration and the resulting area amplification ($f_{\text{IAF}}$) set the geometric preconditions for the mechanical instabilities that drive overlayer folding, which we observe directly as detailed in the following section.

**Facet-induced folding of spill-over grown graphene across Pt grains**

Having elucidated the thermodynamic driver (activity difference) and the dynamics mechanism (surface reconstruction), we now focus on the collective dynamic evolution of CVD graphene growth across polycrystalline Pt grains. By spatiotemporal-resolution time-lapse *in situ* ESEM (**Figure 3a$_1$–a$_6$**), we identify how adjacent grains interact during nucleation and folding.

There were two Pt facets in observation sight, one of which was (14 10 19) and the other was (7 7 26). Continuous observation shows that graphene first nucleates and expands on the highly active Pt(14 10 19) facet. It then spills over across the GB to the adjacent, less active Pt(7 7 26) facet (**Figure 3b,c**). Here, the process documented in the previous section manifests macroscopically: the graphene layer undergoes geometric reconstruction during which graphene folded along certain direction. Specifically, secondary-electron work-function contrast[48] reveals periodic stripe modulations on Pt(7 7 26) arising from graphene-induced step bunching (mid-frame in **Figure 3a$_3$-a$_5$**; schematic atomic model in **Figure 3d**). This reconstruction amplifies the effective contact area ($f_{IAF}$) as defined previously (**Figure 3d**).

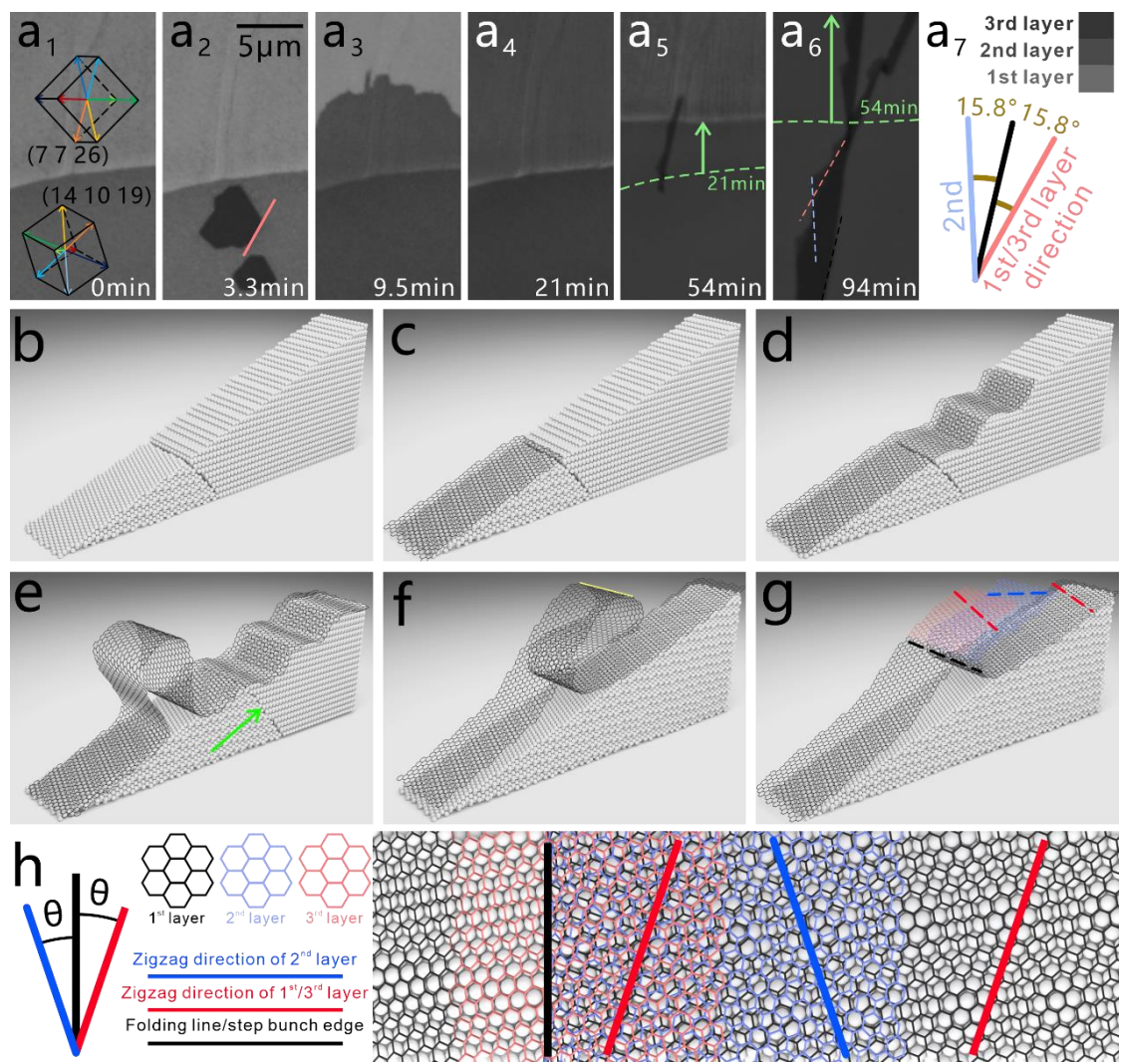

**Figure 3.** Multilayer graphene with defined angles via grain boundary movement. (a1–a6) *In situ* ESEM images illustrating graphene growth, spill-over, wrinkle formation, and tearing processes on (7 7 26) and (14 10 19) grains (1000–1400 °C, 25 Pa, $H_2$:$C_2H_4$ = 100:1). (a7) Schematic diagram of contrast colors for each multilayer graphene layer and the angle between the growth front and the fold line. Inset cubes in (a1) represent Pt surface models for (7 7 26) and (14 10 19). In (a2, a6, a7), red, blue, and black lines denote the zig-zag directions of the 1st/3rd and 2nd graphene layers and the folding line, respectively. (b–g) Schematic representation of graphene wrinkle engineering for multilayer graphene formation. (h) Top view highlighting zig-zag directions of each graphene layer and the fold line/step bunch edge. Zig-zag directions of the 1st and 3rd layers align with the atomic close-packing direction of the (111) plane; the 2nd layer is symmetric to the 1st/3rd layers about the fold line.

As the substrate GB migrates from the (14 10 19) toward the (7 7 26) grain (**Figure 3e**), graphene originally covering the highly reconstructed step-bunched surface is

"transferred" onto the smoother (14 10 19) surface. The resulting mismatch in graphene-Pt contact area forces the excess graphene to detach, driving wrinkle formation along the step-bunch direction (**Figure 3e**, green arrows; **Figure 3f**). Under high temperature and $H_2$-rich conditions, these wrinkles collapse and rupture, creating local adlayer nuclei that subsequently grow into multilayer stacks (**Figure 3f,g**) with a fixed twist configuration (**Figure 3a$_7$,h**)[35]. The third layer AA aligns with the first layer, forming a typical twisted trilayer structure.

These observations reveal that facet-dependent catalytic activity and graphene-induced surface reconstruction act cooperatively to determine the eventual twist angle. However, although we observe that the folding yields a specific angle (*e.g.*, ±15.8° in **Figure 3**), achieving programmable control still requires a precise understanding of what dictates the overlayer in-plane alignment and decides the folding axis. This leads us to the critical role of atomic step-edge orientation.

## Coupled regulation of folding direction by surface step structure and graphene growth orientation

Beyond catalytic reactivity and surface reconstruction, the geometry of atomic step edges provides another lever of control; we therefore analyze how step orientation determines graphene's in-plane alignment and wrinkle direction, thereby providing the structural basis for TGLs with programmable twist angles.

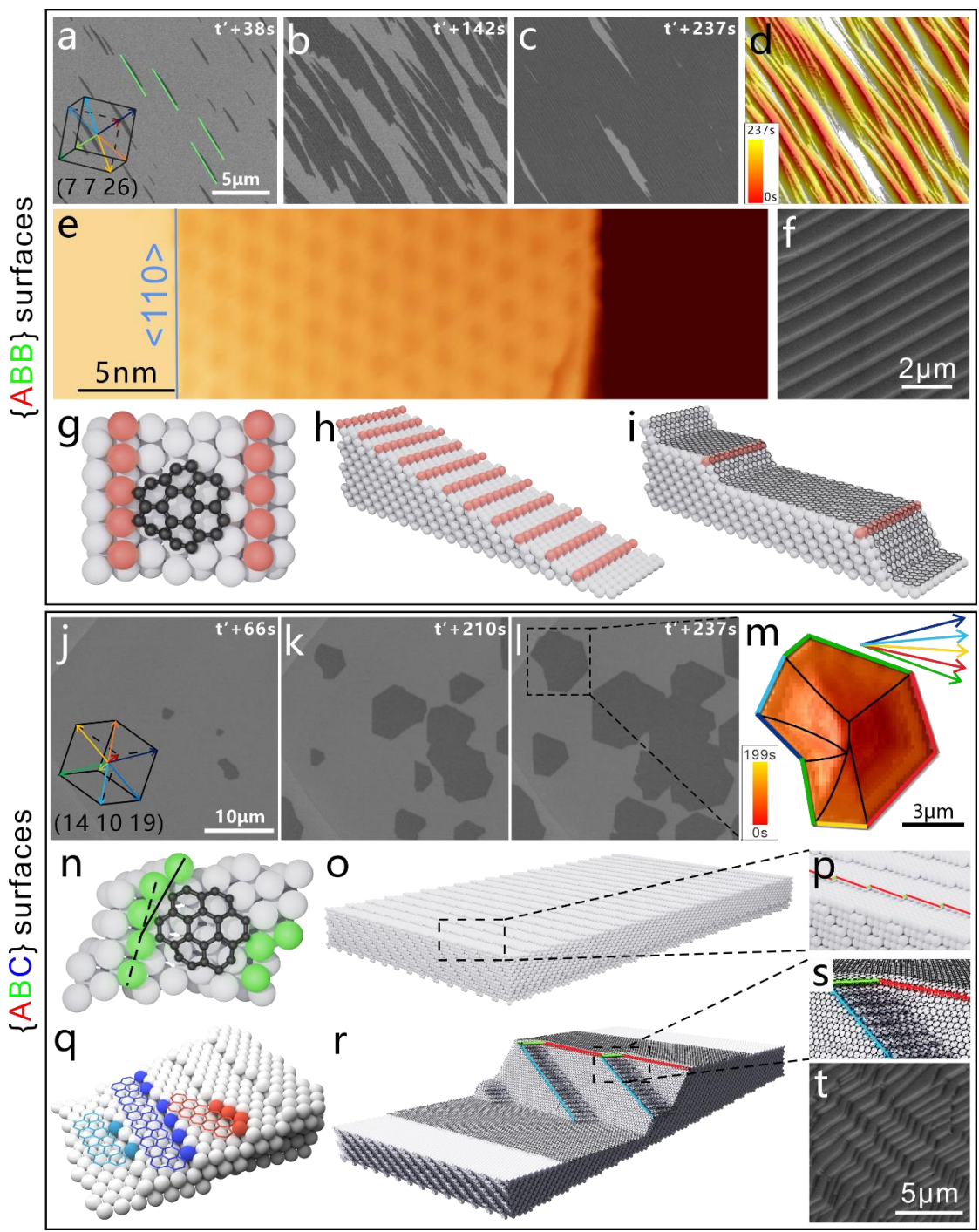

**Figure 4.** Correlation between graphene in-plane orientation, surface structure, and step bunching. (a–c) *In situ* graphene growth on Pt surfaces with close-packed step edges. Inset in (a) shows the (7 7 26) grain orientation; green lines indicate the <110> direction. (d) Time-stacking image of the growth process in (a–c). (e) STM image of graphene

growth along a close-packed step edge; blue line indicates the <110> direction. (f) ESEM image of reconstructed Pt surface with close-packed step edges. (g) Schematic of graphene nucleation at a step edge. (h, i) Models illustrating surface reconstruction of close-packed step edges under graphene coverage. Red spheres in (g–i) highlight <110> direction step edges. (j–l) *In situ* growth of graphene on Pt with kinked steps; inset in (j) shows the (14 10 19) grain model. (m) Time-stacking image capturing growth evolution from t' to t' + 199 s. Colored lines denote distinct zig-zag directions; black lines delineate graphene grain boundaries (GBs). (n) Schematic of graphene nucleation on the Pt(14 10 19) surface with kinked steps. Dashed and solid lines denote close-packed step edges and zig-zag growth directions, respectively; green spheres highlight kinked step edges. (o, p) Surface morphology with kinked steps. (q) Schematic of graphene atomic orientations resulting from kink-atom instability. (r, s) Graphene growth and surface reconstruction; colored lines indicate step and step-bunch edges. (t) ESEM image of the reconstructed Pt surface lacking close-packed step edges.

Direct observations of graphene growth on surfaces with close-packed step edges, specifically the (7 7 26) plane, reveal that graphene domains exhibit a uniform orientation with all growth edges aligned along the <110> direction (highlighted in **Figure 4a** by green lines, **Figure 4a-d**). By integrating STM observations with theoretical simulations, we established a mapping between facet structure and in-plane growth orientation: when the substrate features straight, close-packed <110> steps, the graphene zig-zag edges anchor directly to these step edges (**Figure 4e**). This anchoring markedly suppresses rotational freedom during both nucleation and lateral expansion (**Figure 4g-i**), consequently forcing graphene domains to adopt a single, controllable orientation where the zig-zag direction remains parallel to the step direction (**Figure 4e, g**). In contrast, on surfaces featuring kinked steps, both steps and kinks serve as active nucleation sites that facilitate additional nuclei formation, leading to multiple orientations as revealed by *in situ* ESEM (**Figure 4j-l**); these orientations and the resulting graphene GBs can be directly identified and demarcated via the time-stacking method (**Figure 4m**). Because the energy barrier separating growth along step and kink directions is negligible (The dashed line and solid line in **Figure 4n** indicate the two directions), multi-oriented nucleation inevitably ensues, yielding a polycrystalline overlayer (**Figure 4n, q**). Therefore, to effectively regulate the in-plane orientation of graphene, the growth facet should be maintained kink-free with straight <110> steps.

Moreover, substrate surfaces exhibiting kink–step–terrace (KST) morphology (**Figure 4o–t**) develop irregular facet combination after graphene-induced reconstruction (**Figure 4r–t**). This complex microtopography substantially diminishes control over the subsequent wrinkle direction, impeding predictable synthesis of TGL on such surfaces. Hence, wrinkle-directing facets should satisfy the constraint of kink-free, straight, close-packed steps along <110>. In FCC Pt, {ABB}-type facets (*e.g.*, Pt(7 7 26)) provide straight, close-packed <110> steps; upon graphene coverage and reconstruction, these steps evolve into well-ordered facets comprising two low-index step types with uniform edge character (**Figure 4f, h, i**). As a result, nucleation and wrinkle formation are confined to a single direction parallel to the close-packed steps, and substrate GB migration and step collapse yield a predictable, fixed interlayer twist angle that dictates the adlayers stacking sequence.

Based on the conclusions above, we employ {ABB}-type facets as substrates for producing TGLs with controllable twist angles. During growth, the nucleation facet furnishes the initial orientation template, while a second, wrinkle-directing facet defines the folding axis. By precisely designing the orientations and emergence sequence of these two facets, one can lock both the wrinkle nucleation sites and the interlayer twist angle during growth. In following, we will detail how to regulate the growth order across distinct facets and elucidate the wrinkle formation mechanism driven by GB migrating.

**Controlling the twist angle of TGLs via surface-activity differences and step-structure engineering**

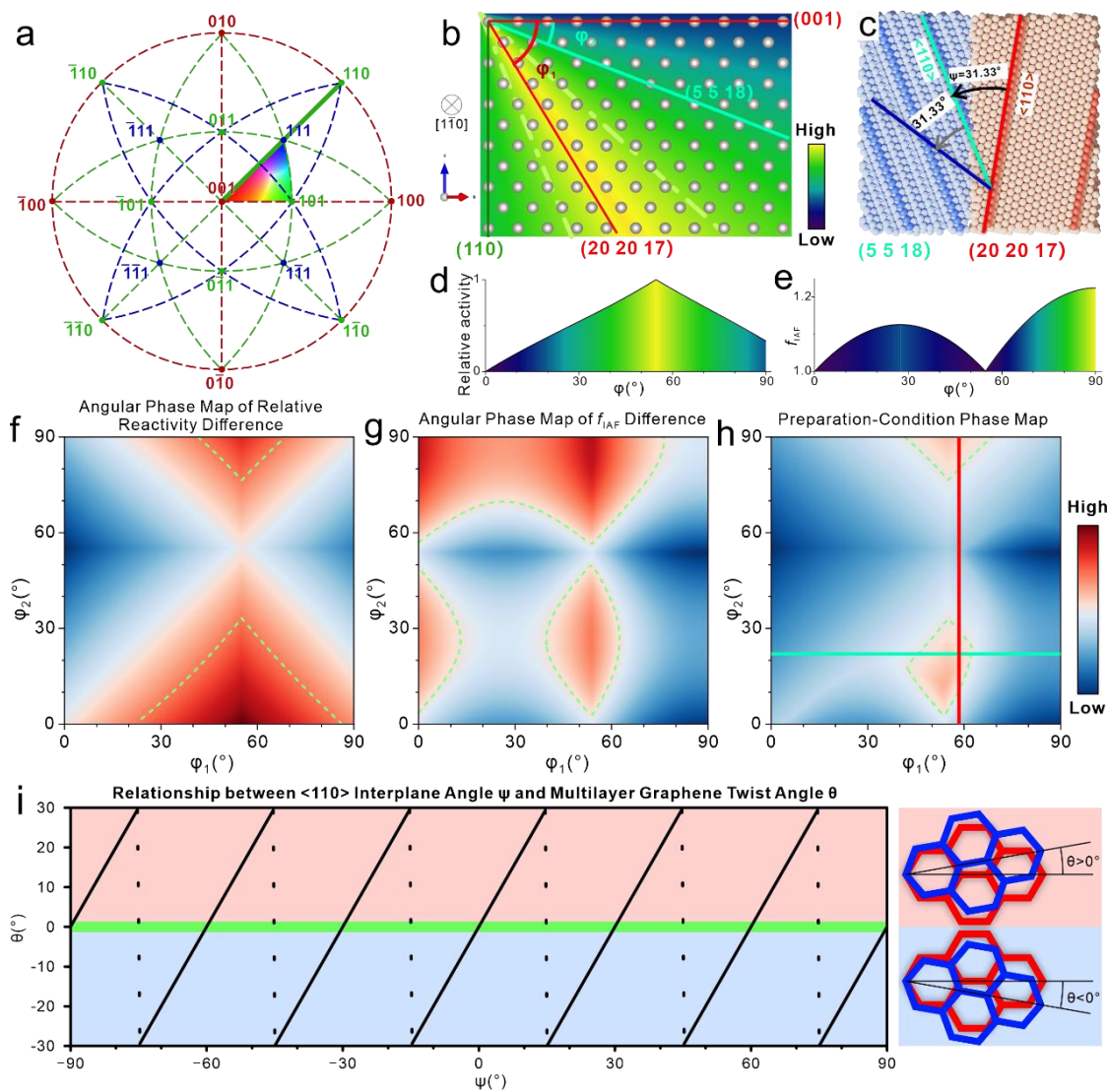

**Figure 5.** Surface structure and growth conditions for controllable-twist multilayer graphene. (a) Standard stereographic projection of Pt; green curves and solid segments indicate symmetry-equivalent representative surfaces with close-packed <110> step edges. (b) Pt atomic model ([1-10] view) parameterized by tilt angle $\varphi$ relative to the (001) plane; color-coded inset indicates relative reactivity. (c) Models of (5 5 18) and (20 20 17) grains; solid lines indicate <110> step and expected zig-zag directions of the 2nd graphene layer; red and cyan markers correspond to (b) and (h). (d, e) Relative reactivity and interfacial area amplification factor ($f_{IAF}$) as functions of tilt angle $\varphi$. (f, g) Differences in relative reactivity and $f_{IAF}$ between target surfaces ($\varphi_1$ and $\varphi_2$) for prescribed twist; the green dashed region indicates the formation window inferred from experiments and simulations. (h) Feasible surface-condition window for controllable-twist multilayer graphene derived from (f) and (g). (i) Relationship between the angle $\psi$ formed by <110> step directions and the resulting graphene twist angle $\theta$. $\theta$ is reported in [−30°, +30°]; $\theta > 0$ and $\theta < 0$ denote counter-clockwise and clockwise chirality, respectively.

Integrating these mechanistic insights, we construct a quantitative design framework that couples facet activity, surface reconstruction, and step geometry to preset controllable interlayer twist angles. As mentioned above, kink-free surfaces bearing close-packed <110> steps are a prerequisite for predictable graphene in-plane orientation and controllable wrinkle formation. Because the surface atomic structure is tied to the Miller indices[49], all surfaces that expose only <110> steps can be represented on the standard stereographic projection of an FCC crystal (**Figure 5a**).[50] In this projection, surfaces have smooth, kink-free step edges if and only if their indices satisfy {ABB}. In the pole figure (**Figure 5a**), the green line delineates the {ABB} family of planes, and all other symmetry-equivalent planes can be mapped equivalently onto this green line[51]. We denote by φ the tilt of a {ABB} facet relative to (001); φ can be measured uniformly in a side-view geometry along $[1\bar{1}0]$ (**Figure 5b**). Hence φ serves as a single parameter to characterize—and stand in for—the {ABB} family.

Therefore, coupled control of oriented nucleation and oriented wrinkling is achieved by pairing two different functional facets (**Figure 5c**). The facet with higher relative activity serves as the nucleation facet. The facet with lower activity, but a larger interfacial area amplification factor ($f_{IAF}$) after reconstruction, serves as the reconstruction facet for graphene folding. Based on the experimental observations that graphene preferentially grows near (111) and the facet-dependent activity established on Pt (**Figure 1k**), we obtain the activity A(φ) within the {ABB} family (**Figure 5d**; color bar), and annotate each φ position in **Figure 5b** with the corresponding color for direct visual cross-reference. To evaluate the contribution of reconstruction to graphene–substrate interfacial area, we approximate the reconstructed surface as a combination of {111}/{100} steps (with {110} unstable and further decomposing into {111}/{100}[42, 44, 52]), thereby deriving $f_{IAF}(φ)$ (**Figure 5e**).

Accordingly, selection of the two facets for oriented nucleation and oriented wrinkling must satisfy three criteria, namely (i) both facets belong to the {ABB} family to ensure straight, kink-free <110> direction steps (**Figure 5a**), (ii) the nucleation facet exhibits higher catalytic activity than the reconstruction facet, and (iii) the reconstruction facet exhibits a higher $f_{IAF}$ than the nucleation facet. Mapping the angles $φ_1$ (nucleation facet) and $φ_2$ (reconstruction facet) into a two-parameter space, we construct the functions of the relative activity difference $ΔA=A(φ_1) − A(φ_2)$ (**Figure 5f**) and the interfacial amplification difference $Δf=f_{IAF}(φ_1) − f_{IAF}(φ_2)$ (**Figure 5g**). Guided by experimental practice, to ensure single-orientation nucleation exclusively on the nucleation facet and directed wrinkle formation on the reconstruction facet, we restrict the usable region to $ΔA≥0.40$ and $Δf≤−0.03$, which is bordered by the green dashed lines in **Figure 5f,g**. Superimposing these boundary conditions yields a preparation-conditions diagram; the area enclosed by the green dashed lines in **Figure 5h** marks the admissible facet-parameter space. Only ($φ_1$, $φ_2$) pairs within this region enable controlled-angle TGL synthesis. Choosing two facets from this area and following the typical CVD workflow, one is supposed to be able to produce TGLs with the preset twist angles.

For coupled control, the in-plane orientation of graphene is locked by the <110> step direction of its initial nucleation facet, while the wrinkle direction (folding axis) is set by the <110> step direction on the reconstruction facet. Thus, by tuning the angle ψ between the <110> step directions of the two facets (**Figure 5c**), one indirectly sets the interlayer twist angle. Specifically, each graphene layer aligns its zig-zag direction with the <110> steps of its respective substrate, giving in **Eq. (2)**

$$\theta = 2\psi - 60°\lfloor\frac{2\psi}{60°} + \frac{1}{2}\rfloor (unit: deg) \qquad (2)$$

where under the six-fold symmetry of the graphene lattice, $\theta$ is restricted to [-30°,30°]. As shown in **Figure 5i**, the target window of inducing magic angle is indicated in green.

As an example (**Figure 5c**), for the (20 20 17) and (5 5 18) facets, the former—used as the nucleation facet—exhibits higher relative reactivity (difference ΔA = 0.518), whereas the latter—used as the reconstruction facet—shows a larger interfacial amplification (difference Δf = −0.070). Setting the angle between their <110> step directions to ψ=31.33° yields an interlayer twist angle of θ = 2.65° from **Eq. (2)**. Together, these parameters define the predictive TGLs growth framework that we implement experimentally in the next section.

# Producing twist angle-controlled TGLs via the facet pairing strategy

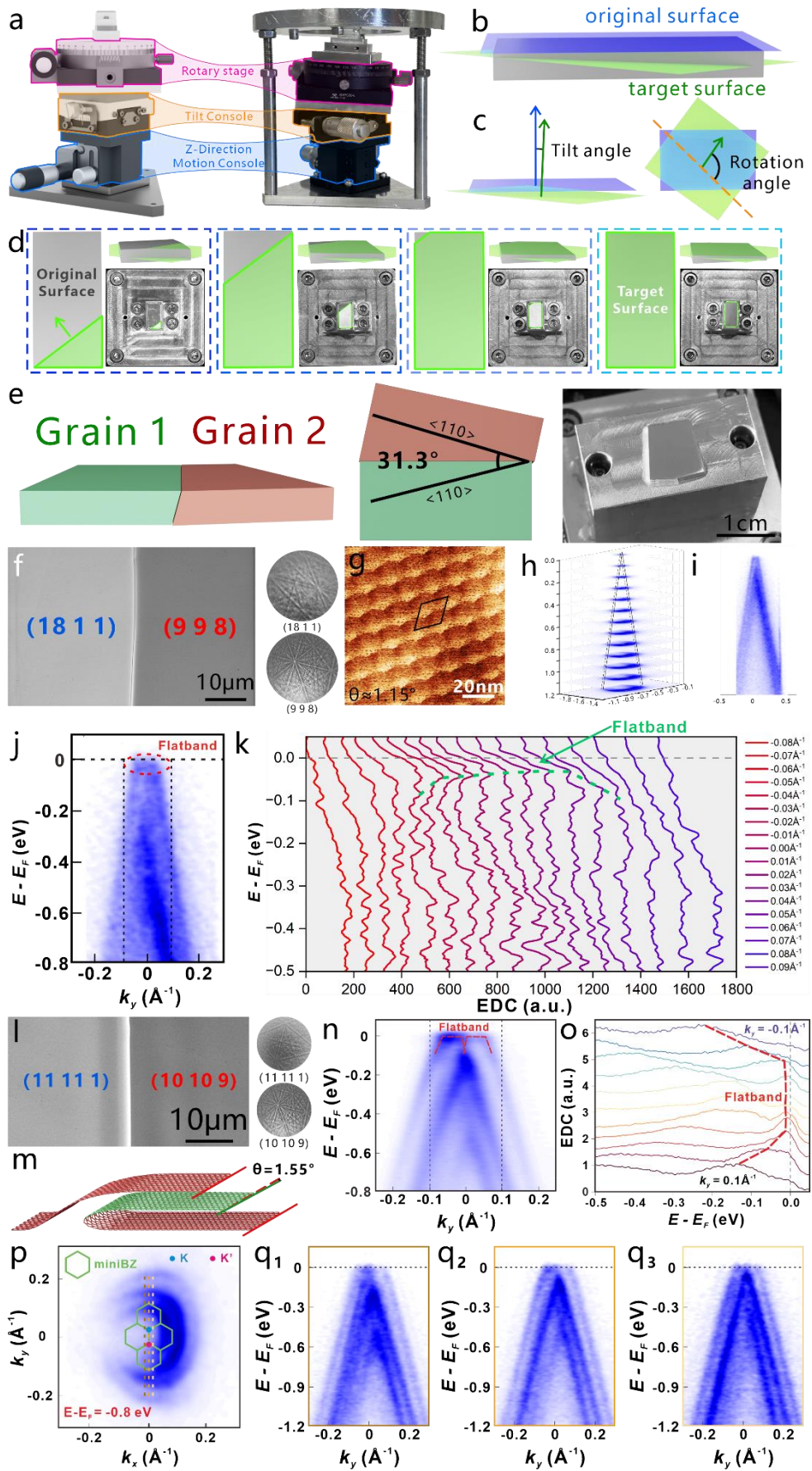

**Figure 6.** Schematic diagram illustrating the method for pre-designed grain grinding and characterization of the electronic structure of multilayer graphene with controlled twist angles and number of layers. (a) Model and physical image of OCSP, a device designed to create a clear crystal surface by tilting, rotating, and bringing the grain into contact with the grinding disc to achieve precise polishing from the original to the target surface. (b) Demonstration of the crystal surface before and after milling. (c) Geometric models of tilting and rotation, corresponding to the tilt console and rotary stage in (a); arrows represent the surface normal vectors. (d) Models of the surface polishing process from the original surface to the target surface of a Pt crystal, with the arrow indicating the expansion direction of the polished surface. (e) Models and physical drawing showing the splicing of two grains with target crystal surfaces at a specific pre-designed angle, followed by thermal treatment for fixation. (f–i) ESEM images of bi-crystal samples (with EBSD Kikuchi patterns), alongside AFM images and ARPES data of the corresponding samples. (g) Diamond-shaped frames in (g) mark the moiré period of graphene on the Pt surface. (h) shows the 3D plot of two measured intersecting Dirac cones of TGL at the *K* point of the graphene Brillouin zone; the energy-momentum dispersion along the marked direction is displayed in (i). (j) Detailed band dispersion of the 1.15° sample, where the red dashed circle highlights the flat band near the Fermi level ($E_F$) with enhanced band weight. (k) Momentum-resolved energy dispersion curves (EDCs) within the interval between the two dashed lines in (j) (marking -0.08 $Å^{-1}$ and 0.09 $Å^{-1}$, respectively). The flat band near $E_F$ is indicated by the green dashed curve, and the precise momentum for each EDC is mapped via the color bar on the right. (l–q) Data for another MATTG sample fabricated by crystal engineering. (l) ESEM image of bi-crystal samples with EBSD Kikuchi patterns. (m) Schematic illustration of the MATTG stacking, consisting of three graphene layers where the top and bottom layers are aligned, and the middle layer is twisted by a magic angle of *θ* = 1.55°. (n) High-resolution ARPES intensity plot showing band dispersion along the $k_y$ direction (as defined in (p)). A prominent, non-dispersive flat band is clearly visible near $E_F$. (o) Corresponding EDCs extracted from the momentum interval defined by the dashed lines in (n), with a momentum step of 0.02 $Å^{-1}$. The dashed curve highlights the flat band position near $E_F$. (p) Constant energy contour taken at a binding energy of *E*–*$E_F$* = -0.8 eV. The overlay shows the reconstructed moiré mini-Brillouin zones (miniBZ, green hexagons) and high-symmetry points *K* and *K'*. (q1–q3) Representative energy-momentum cuts along the dashed lines in (p); the frame colors of the plots correspond to the colored lines in (p).

Finally, we translate this predictive framework into practice by fabricating bi-crystal Pt substrates with programmed facet pairs, experimentally realizing the controllable TGLs synthesis and validating the resulting twisted graphene layers. Here, one facet pair contains two single-crystal surfaces dedicated to oriented nucleation and controllable reconstruction, respectively, and the key factor is the preset angle ψ between their <110> step directions.

To prepare target facets precisely, we developed and used an orientation-controlled surface polisher (OCSP; **Figure 6a**) providing three degrees of freedom—rotation, tilt, and vertical lift. We first indexed the crystallographic and in-plane orientations of the starting surface by EBSD (**Figure 6b**), calculated the required tilt and in-plane rotation to reach the target facet relative to the initial one (**Figure 6c**), and then fixed the crystal on the OCSP to adjust and polish to the desired orientation. The simulated geometry and the fabricated surface show close agreement in exposed-facet morphology and green outline shape (**Figure 6d**), confirming accurate and reproducible orientation control. Using the same workflow, we prepared the nucleation facet and the reconstruction facet, forming the bi-crystal pair. The remaining surfaces were refined to be coplanar, and the <110> step directions of the two target facets were aligned to the preset angle ψ (example ψ ≈ 31.3°; **Figure 6e**).

The two facets were then thermally bonded under 80% $H_2$/20% $O_2$ at 1000 Pa and 900 °C. We examined the GB region by ESEM and verified the orientations by EBSD to ensure the intended facet pairing. During growth, a laser spot was placed right beneath the nucleation facet to establish a temperature gradient from the nucleation facet toward the reconstruction facet, driving the substrate GB to migrate accordingly.

Following the previously established sequence of nucleation, spill-over, and folding, we obtained TGLs with the preset twist angle at the bi-crystal surface. It is worth mentioning that the Frank-van der Merwe growth of graphene on Pt allows for precise control over the number of graphene layers.[53, 54] By calibrating the partial pressures required for graphene formation on specific nucleation facets and those required for spill-over onto reconstruction facets in ESEM, we can achieve full-coverage TGLs with a controlled layer count. Due to the formation of folded graphene wrinkle structures, TGLs can be grown with up to three layers across the entire Pt substrate, forming twisted trilayer structures who has different band dispersion compared with twisted bilayer graphene. Moreover, TGLs layer count tunning is inversible. By manipulating the hydrocarbon partial pressure, topmost adlayers can be selectively etched away[55], leaving the full-coverage bilayer TGLs. Succeeding the growth of TGL, we applied gas-intercalation method to facilitate detachment, transferring the so-grown TGL samples from Pt surface to silicon chip for further characterization.

AFM resolved the moiré superstructure arising from twist. ARPES confirmed the twist angles and the corresponding electronic signatures, including flat-band-related features at magic angle[56-59] $\theta \approx 1.15°$ (**Figure 6i**) and separated Dirac cones at $\theta \approx 2.65°$ and $5°$. For the magical angle twisted bilayer graphene (MATBG), the flat band near $E_F$ is clearly shown in the energy-momentum cut (**Figure 6j**) and demonstrated in detail through the momentum-resolved energy distribution curves (EDCs) (**Figure 6k**). Moreover, through adjusting the hydrocarbon condensation to a proper value, we also synthesized magic angle twisted trilayer graphene (MATTG)[60, 61] with prominent non-dispersive flat band at $E_F$ (Figure 6n,o). This flat band is observable through the whole K-pocket of graphene, reflecting the tunability of moiré potential (Figure 6p,q). Together, these morphological and electronic characterizations validate the feasibility and effectiveness of our strategy, demonstrating the potential of realizing strong correlation physics in so-grown TGL samples. Last but not least, it should also be noted that the producing capacity through our framework can be dramatically increased through our specially engineered surface reconstruction and material platform (SRMP), where we can create the same chemical environment as in the ESEM chamber (connected with SRMP) and therefore realize precise synthesis of multiple TGL samples.

**Conclusion**

In summary, this work begins with Pt facet-dependent catalytic activity and graphene-induced substrate reconstruction, clarifying the intrinsic linkage between nucleation–growth–folding kinetics and the emergence of interlayer twist angles. On this basis, we formulate a predictive, reproducible facet-engineering scheme: pair a high-activity, low interfacial area amplification factor nucleation facet with a low-activity, high interfacial area amplification factor reconstruction facet, and prescribe the interlayer twist via geometric parameters, thereby enabling controlled synthesis of multilayer graphene. Notably, this workflow is built on typical CVD growth, ensuring strong process compatibility and scalability and thus providing a general route to controlled-twist two-dimensional materials under conventional CVD conditions. Importantly, the approach may not be confined to the Pt/graphene system; the paradigm of creating geometry-set angles through facet-dependent reconstruction is broadly applicable. Validation with graphene as a prototype two-dimensional material indicates that the method naturally extends to other metal catalysts and other foldable two-dimensional materials, offering a practical pathway from mechanistic understanding to scalable, controlled manufacturing.

## Methods

### *In situ* ESEM

Experimental investigations were conducted *in situ* within a home-modified ESEM (Thermo Fisher Quattro-S and FEI Quantum 200). This setup features a fully custom-built differential pumping system, enabling the equipment to operate at pressures as high as 0.1 bar. To maintain a clean vacuum environment, oil-free pre-vacuum pumps were employed. The instrument was further enhanced with a bespoke laser-heating assembly and a precise gas supply module utilizing Bronkhorst mass flow controllers; gas phase composition was analyzed using a Pfeiffer HiQuad mass spectrometer. Additionally, the instrument was equipped with an EBSD detector to acquire EBSD patterns. We performed plasma cleaning of the ESEM chamber before each session to eliminate contaminants. The sample size used for the experiments was 3×3 mm. Temperature measurements were facilitated by a B-type thermocouple meticulously spot-welded onto the substrate. The microscope was operated at 2 kV using a home-made detector for image acquisition, and we confirmed that the electron beam exerted no detectable influence on the graphene growth evolution.

### ARPES

After synthesis in the ESEM, the graphene samples were transferred to the ARPES end station using a dedicated vacuum suitcase. These experiments were carried out at the SSRF BL07U beamline, where the chamber base pressure was kept below $1 \times 10^{-10}$ mbar. Prior to data collection, the samples were degassed by annealing at 180 °C for 3 hours to eliminate any potential adsorbates. The subsequent ARPES characterizations were executed at T = 20 K with an incident photon energy of 92.6 eV, providing high-resolution electronic structure information.

**Quasi-*in situ* AFM**

Upon completion of the graphene growth process within the ESEM, the specimens were moved to the AFM system. For topographical imaging, we utilized the tapping mode with a high-performance SNL-10 probe from Bruker. The measurements were specifically conducted using a V-shaped silicon-on-nitride cantilever (designated as Probe B). This probe is characterized by a spring constant ($k$) of approximately 0.32 N/m and an operational resonance frequency ($f_0$) between 40 and 75 kHz, ensuring high-resolution and stable imaging of the graphene surfaces.

**Quasi-*in situ* STM**

Quasi-*in situ* STM characterization was performed using a commercial SPECS JT-STM setup. The experimental workflow involved transferring the graphene/Pt sample from the ESEM growth chamber to the STM system using a vacuum-sealed transfer suitcase, thereby maintaining UHV conditions. Once loaded, the sample was cooled to a base temperature of 5 K on a cryogenic stage for high-resolution imaging. We employed a commercial Pt–Ir tip for both topographic imaging and tunneling spectroscopy; the tip was meticulously calibrated on a reference system consisting of silver islands on p-type Si(111)-7×7 surfaces before the experiments.

**NAP-XPS**

The NAP-XPS characterization during $C_2H_4$ CVD on Pt substrates was conducted at the ISISS end station of the Fritz Haber Institute of the Max Planck Society, located at the BESSY II synchrotron facility. Utilizing a specialized differentially pumped XPS setup, we performed measurements under CVD pressures reaching up to 1 mbar, despite a system base pressure of less than $10^{-7}$ mbar. This configuration enabled the real-time monitoring of surface species throughout the deposition process.

**DFT calculation**

The Vienna ab inito simulation package (VASP) was employed for DFT calculations[62, 63]. The Perdew-Burke-Ernzerhof parameterization of generalized gradient approximation (GGA) was used for the exchange-correlation function[64]. The ion-electron interactions were embodied in projector augmented wave method with a cutoff energy of 400 eV[65, 66]. The Grimme DFT-D3 method was used for the corrections of the interlayer van der Waals interaction[67].

The formation energy of graphene on Pt(111) was calculated using a model consisting of 8×8 graphene on a 7×7, 4-layer thick Pt(111) substrate, within a periodic box. A 15Å thick vacuum slab is introduced to eliminate the image interference perpendicular to the plane. The substrate is slightly compressed to match the supercell lattice of graphene. The formation energy of graphene on Pt(100) was calculated using a model consisting of graphene on a 4-layer thick Pt(100) substrate, within a 19.7Å × 17Å × 30Å periodic box. The formation energy of precursors was calculated using a model consisting of one precursor molecule located at the top, HCP, FCC or bridge site of

Pt(100) and Pt(111). For these cases, DFT calculations are sampled by $1 \times 1 \times 1$ k-mesh using the Monkhorst-Pack method[68]. For geometry relaxation, the force on atoms is converged below $0.01 eV/Å$. Additionally, the electronic self-consistency criterion is set to $10^{-4} eV$. The energy barriers were calculated by using the climbing image nudged elastic band (CI-NEB) method[69] with a force threshold of $0.01 eV/Å$. The energy profiles of graphene growth were calculated using a model consisting of 8 graphene cells along zig-zag direction on 7 cells of Pt <110> steps. The core-level binding energy shift of carbon atoms was analyzed using the Janak–Slater transition state approximation[70].

**Acknowledgements:**

This work was mainly supported by the National Natural Science Foundation of China under grant no. T2525029. The authors wish to express their sincere gratitude to Dr. Raoul Blume from the Fritz Haber Institute of the Max Planck Society for his invaluable support and expertise in the NAP-XPS measurements. We also extend our thanks to Dr. Yi Cui and Dr. Wei Wei at the Suzhou Institute of Nano-Tech and Nano-Bionics, Chinese Academy of Sciences, for their assistance with the STM and LEED characterizations. Zhu-Jun Wang thanks to Prof. Ding-Yuan Wang from Xi'an Polytechnic University for his guidance. His wisdom and integrity have profoundly influenced personal development beyond research.

**Author contributions:**

Z.-J.W. conceived the project. *In situ* ESEM and EBSD experiments and analysis were conducted by C.X., H.Z., and Z.-J.W. Z.-J.W., Z.Z., and L.S. conducted the AFM experiments. M.S. and Z.Z. conducted the ARPES experiments. X.K. and C.X. performed the theoretical analysis and simulations. Z.-J.W., C.X., M.S., Z.Z., Z.Y., H.Z. and L.S. helped with the sample characterizations and data analysis. C.X. wrote the original draft. Z.-J.W. and M.S. reviewed and edited the draft. Important contributions to the interpretation of the results and conception were made by C.X., M.S. and Z.-J.W.. All authors discussed the results and commented on the manuscript.

**Competing interests:**

Authors declare no competing interests.

**Data availability:**

All data are available in the Article.